\documentstyle[12pt,psfig]{article}

\begin{document}
\baselineskip 0.7cm

\newcommand{\gsim}{ \mathop{}_{\textstyle \sim}^{\textstyle >} }
\newcommand{\lsim}{ \mathop{}_{\textstyle \sim}^{\textstyle <} }
\newcommand{\vev}[1]{ \left\langle {#1} \right\rangle }
\newcommand{\lsp}{ \left ( }
\newcommand{\rsp}{ \right ) }
\newcommand{\lmp}{ \left \{ }
\newcommand{\rmp}{ \right \} }
\newcommand{\llp}{ \left [ }
\newcommand{\rlp}{ \right ] }
\newcommand{\labs}{ \left | }
\newcommand{\rabs}{ \right | }
\newcommand{\EV} { {\rm eV} }
\newcommand{\KEV}{ {\rm keV} }
\newcommand{\MEV}{ {\rm MeV} }
\newcommand{\GEV}{ {\rm GeV} }
\newcommand{\TEV}{ {\rm TeV} }
\newcommand{\YR}{ {\rm yr} }
\newcommand{\mgut}{M_{GUT}}
\newcommand{\mint}{M_{I}}
\newcommand{\mgra}{M_{3/2}}
\newcommand{\mll}{m_{\tilde{l}L}^{2}}
\newcommand{\mdr}{m_{\tilde{d}R}^{2}}
\newcommand{\mllXX}[1]{m_{\tilde{l}L , {#1}}^{2}}
\newcommand{\mdrXX}[1]{m_{\tilde{d}R , {#1}}^{2}}
\newcommand{\mgy}{m_{G1}}
\newcommand{\mgl}{m_{G2}}
\newcommand{\mgc}{m_{G3}}
\newcommand{\nuR}{\nu_{R}}
\newcommand{\slL}{\tilde{l}_{L}}
\newcommand{\slLi}{\tilde{l}_{Li}}
\newcommand{\sdR}{\tilde{d}_{R}}
\newcommand{\sdRi}{\tilde{d}_{Ri}}
\newcommand{\e}{{\rm e}}
\renewcommand{\thefootnote}{\fnsymbol{footnote}}
\setcounter{footnote}{1}

\makeatletter
%
%
%
%
%
\newtoks\@stequation

\def\subequations{\refstepcounter{equation}%
  \edef\@savedequation{\the\c@equation}%
  \@stequation=\expandafter{\theequation}
  \edef\@savedtheequation{\the\@stequation}
  \edef\oldtheequation{\theequation}%
  \setcounter{equation}{0}%
  \def\theequation{\oldtheequation\alph{equation}}}

\def\endsubequations{%
  \ifnum\c@equation < 2 \@warning{Only \the\c@equation\space subequation
    used in equation \@savedequation}\fi
  \setcounter{equation}{\@savedequation}%
  \@stequation=\expandafter{\@savedtheequation}%
  \edef\theequation{\the\@stequation}%
  \global\@ignoretrue}


\def\eqnarray{\stepcounter{equation}\let\@currentlabel\theequation
\global\@eqnswtrue\m@th
\global\@eqcnt\z@\tabskip\@centering\let\\\@eqncr
$$\halign to\displaywidth\bgroup\@eqnsel\hskip\@centering
     $\displaystyle\tabskip\z@{##}$&\global\@eqcnt\@ne
      \hfil$\;{##}\;$\hfil
     &\global\@eqcnt\tw@ $\displaystyle\tabskip\z@{##}$\hfil
   \tabskip\@centering&\llap{##}\tabskip\z@\cr}

\makeatother


\begin{titlepage}

\begin{flushright}
UT-953
\end{flushright}

\vskip 0.35cm
\begin{center}
{\large \bf Right-handed Neutrinos as Superheavy Dark Matter}
\vskip 1.2cm
Yosuke Uehara

\vskip 0.4cm

{\it Department of Physics, University of Tokyo, 
         Tokyo 113-0033, Japan}\\
\vskip 1.5cm

\abstract{We propose that right-handed neutrinos are 
very long-lived dark matter. The long lifetime is realized by 
the separation of the wavefunction of right-handed
neutrinos and that of other fermions in an extra dimension. 
Such long-lived and superheavy dark matter can 
naturally explain observed 
ultra high energy cosmic rays above the GZK cutoff
($5 \times 10^{19} \EV$) and huge amounts 
of cold dark matter simultaneously.
Furthermore, the exponentially suppressed 
Yukawa couplings of right-handed neutrinos 
leads to the high predictablilty on the mass parameter of the
neutrinoless double beta decay, as all the models which
predict very small neutrino mass of one generation.}

\end{center}
\end{titlepage}

\renewcommand{\thefootnote}{\arabic{footnote}}
\setcounter{footnote}{0}

%
%
%
%

\section{Introduction}

One of the long-standing mysteries in modern cosmology is the nature of
dark matter. The energy density of dark matter should be
about 30 \% of that of the universe, but baryons and observed
astrophysical objects can constitute only a few percent of the energy.
Therefore physics beyond standard model is needed to explain 
huge amounts of dark matter.
They may be supersymmetric particles, topological defects or 
composite objects.

Another mystery of cosmology is the existence of ultra high energy
cosmic rays(UHECR). Cosmic rays above the GZK cutoff\cite{GZK}
($5 \times 10^{19} \EV$) cannot be explained by usual 
acceleration mechanisms. 

These mysteries can be explained simultaneously by introducing
long-lived superheavy dark matter $X$\cite{CKR,KR,FKN,BKV}. 
Its lifetime is longer
than the age of the universe, and its mass is larger than
the energy scale of the GZK cutoff. Such particles
would overclose the universe if once they were in a 
thermal equilibrium\cite{GK}. But they can be generated gravitationally
during the reheating epoch just after 
the end of inflation\cite{CKR,KT}.

In this paper,
we identify $X$ as one kind of right-handed neutrino $\nu_{R}$.
For simplicity of discussion, we assume that this $X$ is
electron right-handed neutrino $(\nu_{R})_{e}$, but any generation
of right-handed neutrino can play a role of superheavy dark matter.

The typical mass scale of the right-handed neutrinos is very large, 
$M_{\nu_{R}} \sim 10^{14} \GEV$, in order to
explain the very small mass scale of left-handed neutrinos
by seesaw mechanism\cite{SEESAW}. Thus right-handed neutrinos 
can be candidates of superheavy dark matter. \footnote{in reference
\cite{RHDM}, It is argued that right-handed can be the
candidate of dark matter, but not superheavy.}

We have to explain why the lifetime of such heavy
objects is very long. We can realize the very long lifetime 
by introducing an extra dimension. If the compactification
scale of the extra dimension is extremely small,
no laboratory and astrophysical constraints apply,
so we can use an extra dimension.

We assume that the wavefunction of right-handed 
electron neutrinos localize far away from the localization place of 
other fermions. The interaction between them is exponentially suppressed
because of this separation, and thus we can achieve the
extremely long lifetime.

Furthermore, because of this suppression, 
the mass of the lightest left-handed neutrino
becomes extremely small. This leads to the high predictability 
on the mass scale relevant to the neutrinoless 
double beta decay\cite{DOUBLEBETA}, as all models that predict
very small neutrino mass of one generation.
If the large angle MSW solution\cite{MSW} of 
solar neutrino deficit is correct, 
as recent experiments suggest\cite{SOL}, GENIUS\cite{GENIUS} 
may be able to detect the signal of the neutrinoless double beta decay.

\section{The Model}

We realize the exponentially small Yukawa couplings of right-handed electron
neutrinos by the separation of the wavefunction of the right-handed
electron neutrinos and that of other fermions in the extra dimension. 
Again note that, the Yukawa couplings of electron right-handed 
neutrino do not have to be exponentially suppressed.
It can be muon right-handed neutrino, or tau right-handed neutrino.
But for simplicity, we assume that it is electron right-handed neutrino.

We assume that our world has one extra dimension.
Let us denote the coordinate of the extra dimension by $y$.
Then by using a five-dimensional domain wall scalar field,
we can localize four-dimensional chiral fermions at different
four-dimensional slices in the fifth dimension\cite{AS}. Its wavefunction
$\phi (y)$ is given by
\begin{eqnarray}
\phi (y) = \frac{\sqrt{\mu}}{(\pi/2)^{1/4}} e^{- \mu^{2} (y-y_{0})^{2}},
\end{eqnarray}
where $\mu$ is the mass parameter of the domain wall scalar field,
and $y_{0}$ is the localization position of chiral fermions.
We assume that the right-handed electron neutrinos are localized at $y=r$,
higgs doublets are not localized anywhere, 
and all other fermions are localized 
at $y=0$. It is shown in figure \ref{gaussian}.

\begin{figure}
\centerline{\psfig{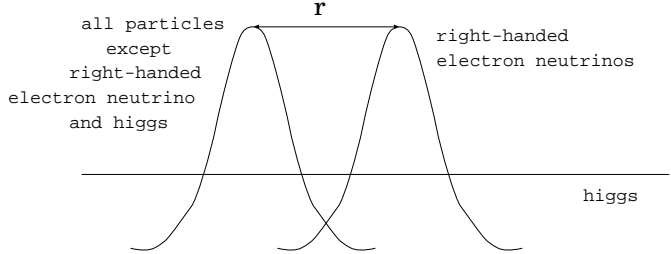}}
\caption{The five-dimensional wave function. The overlap between
right-handed electron neutrinos and other fermions is exponentially
suppressed.}
\label{gaussian}
\end{figure}
 
Thus the overlap between the wavefunction of right-handed 
electron neutrinos and that of other lepton doublets 
is exponentially suppressed and Yukawa coupling terms become
\begin{subequations}
\begin{eqnarray}
{\cal L_{\rm 4d}} &=& \sum_{\alpha} y_{\alpha e} (\bar{l}_{L})_{\alpha} H (\nu_{R})_{e} \int dy \phi_{l_{L}} (y) \phi_{\nu_{R}} (y) \\
&=& \sum_{\alpha} e^{- \mu^{2} r^{2} / 2} y_{\alpha e} (\bar{l}_{L})_{\alpha} H (\nu_{R})_{e}.
\end{eqnarray}
\end{subequations}
Thus Yukawa couplings are exponentially suppressed. Let us denote
all exponentially suppressed quantities by $\epsilon$. Then 
the Lagrangian of neutrino sector becomes
\begin{eqnarray}
{\cal L} = \epsilon_{\alpha e} (\bar{l}_{L})_{\alpha} H (\nu_{R})_{e} + \sum_{\alpha , \beta \neq e} y_{\alpha \beta} (\bar{l}_{L})_{\alpha} H (\nu_{R})_{\beta} + \sum_{\alpha \beta} (M_{\nu_{R}})_{\alpha \beta} (\nu_{R})_{\alpha}^{c} (\nu_{R})_{\beta},
\end{eqnarray}
where
\begin{eqnarray}
(M_{\nu_{R}})_{\alpha \beta} \sim
M \pmatrix{ 
a & \epsilon & \epsilon \cr
\epsilon & b & c \cr
\epsilon & c & d \cr
}.
\end{eqnarray}
Since right-handed mu and tau neutrinos are localized far away
from right-handed electron neutrinos, as other usual fermions,
some parts of Majorana mass matrix of right-handed neutrinos
are also exponentially suppressed. 
By using seesaw mechanism\cite{SEESAW}, 
we can naturally realize the very small left-handed
mu and tau neutrino masses if we take
$M \sim 10^{14} \GEV$ and $b,c,d \sim {\rm O}(1)$.
Since Yukawa couplings of a right-handed electron neutrino are
extremely small, $a$ do not need to be ${\rm O}(1)$, 
but we take $a \sim {\rm O}(1)$ here.

Therefore the mass of the lightest neutrino $m_{1}$ 
is extremely small because of the exponentially small Yukawa coupling:
\begin{subequations}
\begin{eqnarray}
m_{1} &\sim& \frac{e^{- \mu^{2} r^{2}} y^{2} v^{2}}{M} \\
&\sim& e^{- \mu^{2} r^{2}} 10^{-1} \EV.
\end{eqnarray}
\end{subequations}
Here $v$ is the VEV of higgs bosons.
The masses of other neutrinos are determined
by the experiments of atmospheric\cite{ATM} and solar\cite{SOL} 
neutrino oscillations:
$m_{2} \sim \sqrt{\delta m_{{\rm sol}}^{2}}$ and 
$m_{3} \sim \sqrt{\delta m_{{\rm atm}}^{2}}$.
 
\section{Ultra High Energy Cosmic Ray}

It was pointed out that the flux of cosmic rays should be suddenly dumped
at the GZK cutoff ($5 \times 10^{19} \EV$)\cite{GZK} because of the
interaction between cosmic rays and cosmic microwave background
radiation photons. Though the absence of the GZK cutoff is now experimentally
established\cite{AGASA,FLYSEYE,HAVERAH,SUGAR}. So we should propose
a scenario to explain the observed UHECR.

``Top-Down'' scenarios assume some superheavy objects 
with very long lifetime. It may be topological defects\cite{BR,H,BV},
composite objects\cite{BS,HINY,S}, 
or superheavy dark matter\cite{HU,HNY,KR,FKN,BKV}.
(for a detailed review, see \cite{BS_REVIEW}.)
In order to explain UHECR by superheavy dark matter $X$, it must satisfy
the following conditions\cite{KR,BS_REVIEW}.
\begin{subequations}
\begin{eqnarray}
& m_{X}^{} & \gsim  10^{12} \GEV, \\
10^{10} \YR \lsim & \tau_{X}^{} & \lsim 10^{22} \YR, \\
10^{-12} \lsim & \Omega_{X} h^{2} & \lsim 1.
\end{eqnarray}
\end{subequations}

In our scenario, right-handed electron neutrinos are the superheavy
dark matter. Its mass $M_{\nu_{R}} \sim 10^{14} \GEV$ 
naturally leads to the energy above the GZK cutoff. 
The long lifetime is realized by the
exponentially suppressed Yukawa couplings.

In order to explain UHECR and the missing energy of the universe 
simultaneously, we took $m_{\nu_{R}} = 2 \times 10^{14} \GEV$, 
$\tau_{\nu_{R}} =10^{20} {\rm yr}$ and 
$n_{\nu_{R}} = 4.5 \times 10^{-15} {\rm m}^{-3}$. 
This long lifetime can be realized by a moderate fine-tuning:
$\mu r \sim 12$. We can explain the energy of the CDM: 
$\Omega_{\nu_{R}} = 0.2$. This compactification scale 
$r \sim \frac{12}{\mu}$ can be extremely small if the
mass scale $\mu$ is extremely large, so no laboratory and
astrophysical constraints for extra dimensions apply in this case.

And under these conditions,the predicted UHECR flux becomes consistent
with observations as shown in figure \ref{cosmicray}.

\begin{figure}
\centerline{\psfig{file=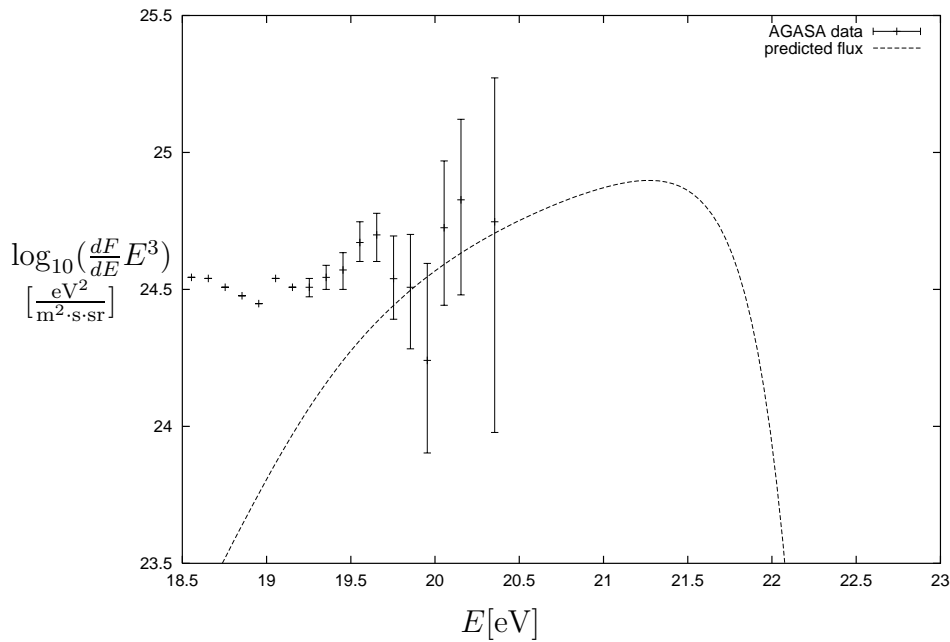,height=8cm}}
\begin{picture}(0,0)
\put(10,140){$\log_{10}(\frac{dF}{dE} E^{3})$}  
\put(15,123){$[\frac{\EV^{2}}{{\rm m}^{2} \cdot {\rm s} \cdot {\rm sr}}]$}  
\put(180,1){$E [\EV]$}
\end{picture}\caption{The observed UHECR events and the prediction of our model.}
\label{cosmicray}
\end{figure}

We calculated this flux by the MLLA approximation\cite{MLLA} and used SUSY QCD.
Though it was pointed out that this is not reliable for large $x$ 
and its normalization is uncertain\cite{BS}, it can be used
as a benchmark.

Since these UHECR mainly come from our galaxy halo because dark matter
is concentrated on it, there should be some anisotropy 
in the observed UHECR events\cite{AGASA_ANISOTROPY}.
Future observatories may be able to see it.

Right-handed electron neutrinos decay into higgs bosons
 and left-handed lepton doublets.
There exists a constraint for heavy particles $X$ 
which can decay into left-handed neutrinos\cite{EGLNS,GGS,Z_BURST}:
\begin{eqnarray}
\frac{\tau_{X}}{t_{0}} BR(X \rightarrow \nu \ {\rm anything})^{-1} 
\ge 2.4 \times 10^{5} (\frac{\Omega_{X}}{0.2})(\frac{h}{0.65})^{2}
(\frac{10^{14} \GEV}{m_{X}})^{3/4}.
\end{eqnarray}
Our model satisfies this condition.

\section{Neutrinoless Double Beta Decay}

If observed, the neutrinoless double beta decay($0 \nu \beta \beta$) 
may become the strongest evidence for the lepton flavor violation. The 
$0 \nu \beta \beta$ decay rate is determined by the mass parameter
\begin{eqnarray}
|m_{\nu_{e} \nu_{e}}| \equiv | \sum_{i} U_{ei}^{2} m_{i} |,
\end{eqnarray}
where $U_{ei}$ is the MNS matrix\cite{MNS}. In our model, the mass of the
lightest neutrino is extremely small,
so its contribution can be neglected. 
As all models that predict very small neutrino mass of one generation,
this fact leads into the high predictability 
of neutrinoless double beta decay.

The mass parameter is described by $U_{e3}$ and 
the parameters of atmospheric and solar neutrino
oscillations\cite{FHY}: $\delta m_{{\rm atm}}^{2} \sim m_{\nu_{3}}^{2}$, 
$\delta m_{{\rm sol}}^{2} \sim m_{\nu_{2}}^{2}$, and 
$\tan^{2} \theta_{{\rm sol}} \equiv |\frac{U_{e2}}{U_{e1}}|^{2}$. It becomes
\begin{subequations}
\begin{eqnarray}
|m_{\nu_{e} \nu_{e}}| &=& | U_{e2}^{2} m_{2} + U_{e3}^{2} m_{3} | \\
&=&|(1-|U_{e3}|^{2}) \sin^{2} \theta_{{\rm sol}} \sqrt{\delta m_{{\rm sol}}^{2}} + |U_{e3}|^{2} e^{i \alpha} \sqrt{\delta m_{{\rm atm}}^{2}}|,
\end{eqnarray}
\end{subequations}
where $\alpha$ denotes the relative phase between the two terms.

We calculate the value of $|m_{\nu_{e} \nu_{e}}|$ for the large angle
MSW solution, the small angle MSW solution, and the LOW solutions.
We take $\delta m_{{\rm atm}}^2 \sim 3,2 \times 10^{-3} \EV^2$.
We also required $|U_{e3}| \le 0.15$ from the CHOOZ experiment\cite{CHOOZ}.

The result is shown in figure \ref{large}, \ref{small} and 
\ref{LOW}. In the case of the large angle MSW solution, $|m_{\nu_{e}
\nu_{e}}|$ is mainly sensitive to the parameter of the solar neutrino
oscillation, $\sin^2 \theta_{{\rm sol}} \sqrt{\delta m_{{\rm sol}}^2}$.
It is very encouraging to see that GENIUS\cite{GENIUS}
may be able to detect the signal of the neutrinoless double
beta decay in most of the paramater region if the large angle MSW
solution, which seems to be the most favored one\cite{SNO_AN}, 
is correct.

\begin{figure}
\centerline{\psfig{file=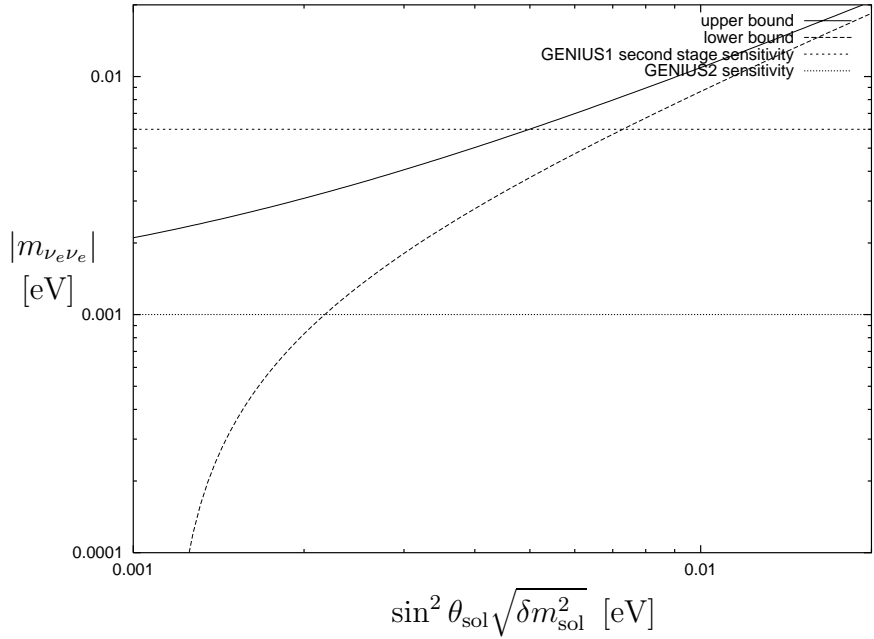,height=8cm}}
\begin{picture}(0,0)
\put(35,140){$|m_{\nu_e \nu_e}|$}  
\put(40,123){$[$eV$]$}  
\put(180,1){$\sin^2\theta_{\rm {\rm sol}} \sqrt{\delta m_{\rm {\rm sol}}^2}\,\,\,[$eV$]$}
\end{picture}
\caption{The predicted value of $|m_{\nu_{e} \nu_{e}}|$ for the large 
angle MSW solution.}
\label{large}
\end{figure}
\begin{figure}
\centerline{\psfig{file=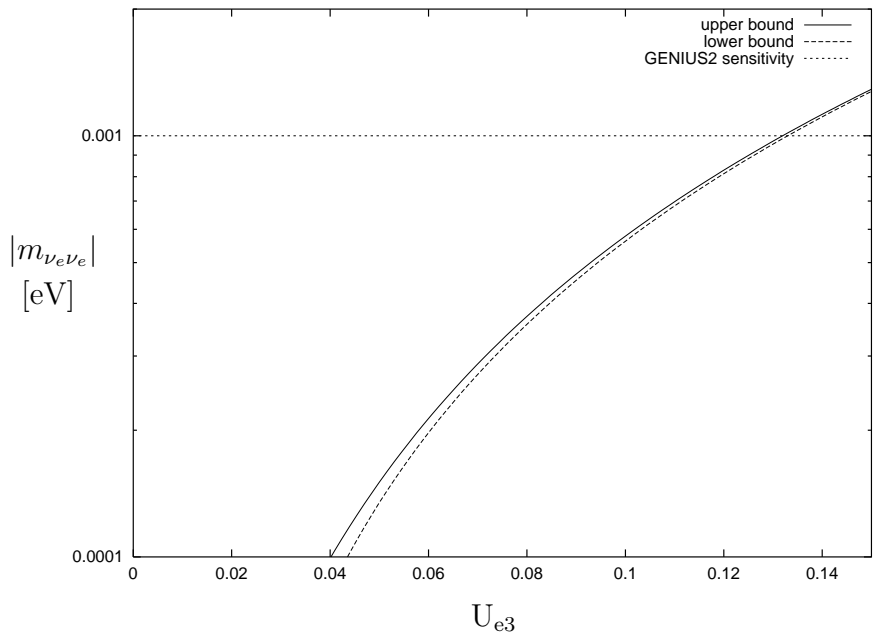,height=8cm}}
\begin{picture}(0,0)
\put(35,140){$|m_{\nu_e \nu_e}|$}  
\put(40,123){$[$eV$]$}  
\put(210,1){${\rm U_{e3}}$}
\end{picture}\caption{The predicted value of $|m_{\nu_{e} \nu_{e}}|$ for the small angle MSW solution.}
\label{small}
\end{figure}
\begin{figure}
\centerline{\psfig{file=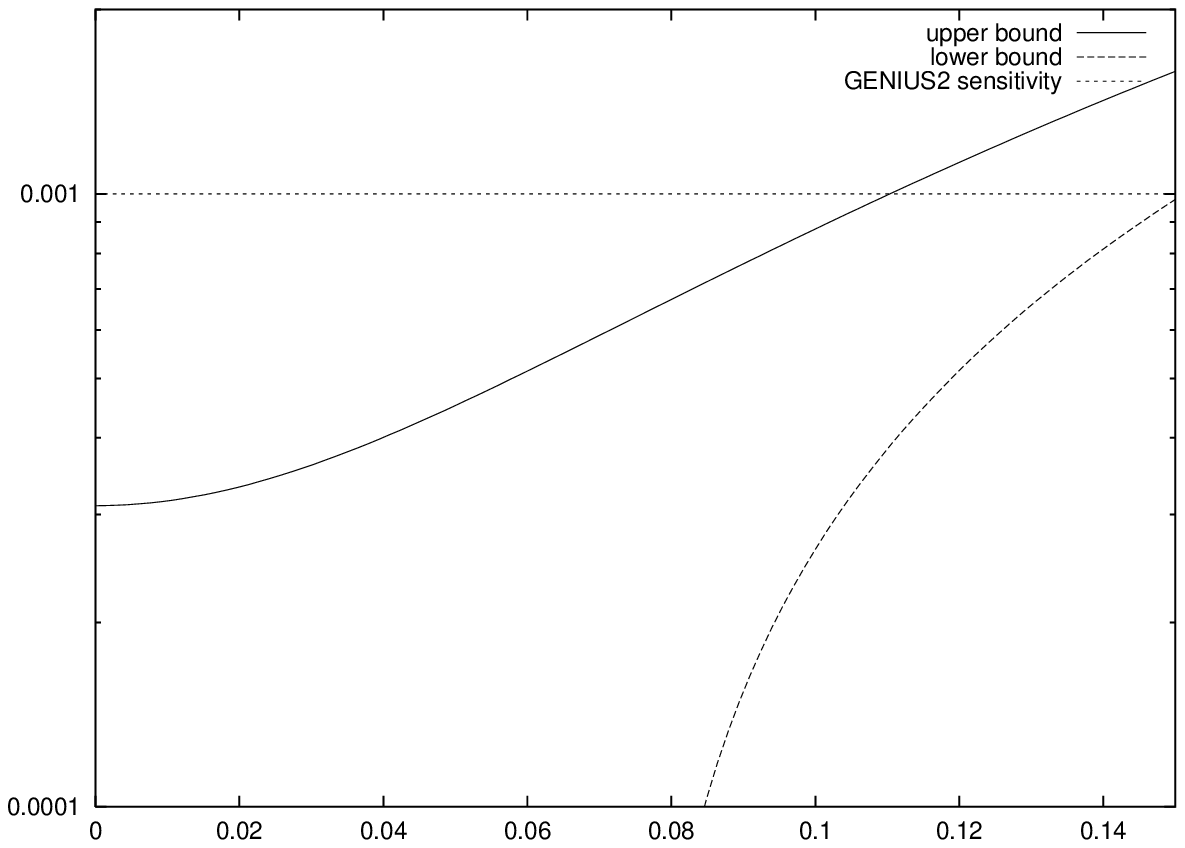,height=8cm}}
\begin{picture}(0,0)
\put(35,140){$|m_{\nu_e \nu_e}|$}  
\put(40,123){$[$eV$]$}  
\put(210,1){${\rm U_{e3}}$}
\end{picture}
\caption{The predicted value of $|m_{\nu_{e} \nu_{e}}|$ for the LOW solution.}
\label{LOW}
\end{figure}

\section{Summary}

In this paper we propose that the right-handed neutrinos
can be candidates of superheavy dark matter, which is needed
to explain huge amounts of cold dark matter 
and UHECR simultaneously. The long lifetime
of the right-handed neutrinos is realized by the separation
between the wavefunction of the right-handed neutrinos and
that of other fermions in the extra dimension. 
Our model also have high predictability on the mass parameter
of the neutrinoless double beta decay.

\noindent
{\bf Acknowledgment}

We thank to M.Fujii, K.Hamaguchi, S.Ryu, H.Takayanagi and 
T.Yanagida for stimulating discussions.

%
%
\newcommand{\Journal}[4]{{\sl #1} {\bf #2} {(#3)} {#4}}
\newcommand{\PL}{\sl Phys. Lett.}
\newcommand{\PR}{\sl Phys. Rev.}
\newcommand{\PRL}{\sl Phys. Rev. Lett.}
\newcommand{\NP}{\sl Nucl. Phys.}
\newcommand{\ZP}{\sl Z. Phys.}
\newcommand{\PTP}{\sl Prog. Theor. Phys.}
\newcommand{\NC}{\sl Nuovo Cimento}
\newcommand{\PAN}{\sl Phys.Atom.Nucl.}

\end{document}